\documentclass[12pt]{article}
\usepackage[utf8]{inputenc}
\usepackage[margin=1in]{geometry}

\usepackage{chngpage}

\title{A U-statistic-based test of treatment effect heterogeneity}
\author{Maozhu Dai, Hal S. Stern\\
    Department of Statistics, University of California - Irvine\\
    Irvine, California, USA
}

\usepackage{etoolbox}
\usepackage{enumitem}
\usepackage{graphicx}
\usepackage{adjustbox}
\usepackage{float}
\usepackage{setspace}
\usepackage{amsmath}
\usepackage{subfigure}
\usepackage{amssymb}
\usepackage{amsthm}
\usepackage{float}
\usepackage{rotating}
\usepackage{blkarray}
\usepackage{longtable}
\usepackage{caption}
\usepackage{siunitx}
\usepackage{mathrsfs}
\usepackage[noadjust]{cite}
\usepackage{geometry}
\usepackage{algorithm}
\usepackage{xcolor}
\usepackage{comment}

\usepackage{natbib}
\setcitestyle{authoryear,open={(},close={)}}

\theoremstyle{plain}

\begin{document}

\maketitle

\section*{Abstract}

Many studies include a goal of determining whether there is treatment effect heterogeneity across different subpopulations. In this paper, we propose a U-statistic-based non-parametric test of the null hypothesis that the treatment effects are identical in different subgroups. The proposed test provides more power than the standard parametric test when the underlying distribution assumptions of the latter are violated. We apply the method to data from an economic study of program effectiveness and find that there is treatment effect heterogeneity in different subpopulations.

\section{Introduction}
Treatment effect heterogeneity is of great importance, as the average treatment effect across the whole population may neglect important variability of the treatment effect across subpopulations. In health care, the concept of personalized medicine is attracting a great deal of attention \citep{jain2009textbook,chan2011personalized,ginsburg2009genomic}, as it promises a way to provide treatment recommendations with greater precision based on a patient's baseline characteristics. In social sciences, people are using similar approaches to assess localized effects in order to comprehensively evaluate a policy or a campaign strategy \citep{bitler2006mean, feller2009beyond}. In one criminology study, arrest can result in effects of opposite sign on recidivism rates for different kinds of  criminals \citep{na2015importance}. In all areas, subgroup analysis \citep{dixon1991bayesian} can be conducted to assess treatment effects at the subpopulation level.
Both randomized studies and observational studies can provide insight into defining relevant subpopulations. In randomized clinical trials, patients can be assigned to different subgroups based on one or several baseline factors.  In observational studies, one approach to identify subpopulation is through subclassification on propensity scores \citep{xie2012estimating}. In either case, reliable inference about whether there is heterogeneity in treatment effects is needed before undertaking an approach based on subpopulations.

There is a great deal of literature on exploring heterogeneity of treatment effects. The published results focus on different aspects of heterogeneity in that they examine different null hypotheses. Some focus on testing conditional stochastic dominance \citep{delgado2013conditional, hsu2017consistent}. Other focus on testing whether the treatment effects in subpopulations are of the same sign \citep{gail1985testing, chang2015nonparametric}. \cite{crump2008nonparametric} and \cite{chang2015nonparametric} are interested in tesing whether the treatment effects are consistently equal to  zero. \cite{ding2016randomization} focus on the null hypothesis that all individuals have the same treatment effects. In this paper, we focus on testing whether the average stratum-specific treatment effects are constant across different strata.

A common approach to identifying whether the stratum-specific average treatment effects are equal across different strata is through parametric statistical tests. \cite{gail1985testing} presented the likelihood ratio test (LRT) assuming the stratum treatment effect estimates follow normal distributions. When there are confounders to be controlled for, regression methods can be applied and treatment effect heterogeneity is tested by testing the interaction term between treatment assignment and effect modifiers  \citep{krishnan2003smoking}. These approaches are widely used but rely on the model assumptions being correct. When the assumptions are not correct, the inference is invalid or the power of the test is impacted. 

Non-parametric approaches exist as well. \cite{crump2008nonparametric} created a non-parametric approach based on a particular 
series estimator for treatment effect introduced by \cite{imbens2005mean}. \cite{sant2020nonparametric} generalized \cite{crump2008nonparametric} by allowing censored data and endogenous treatment selections. 
In this paper, we propose a U-statistic-based approach to test whether the stratum treatment effects are homogeneous without having to estimate the stratum treatment effects.  Our approach relies on an unconfoundedness assumption in each stratum.

U-statistics have been widely used to create distribution-free tests. Examples include the signed rank test and Mann-Whitney test \citep{van2000asymptotic}. Compared to their parametric counterparts assuming normal distributions, U-statistic-based tests usually have higher power when the distributions are far from normal. When the normality assumption is satisfied, the parametric test has slightly higher power, but when the U-statistic-based test is more powerful, the advantage can be significant \citep{hodges1956efficiency, zimmerman1998invalidation, lehmann1975nonparametrics}. The non-parametric heterogeneity test we propose here is also based on U-statistics. We use U-statistics to compute a test statistic comparing treatment effects across pairs of strata. The overall test statistic is a combination of the pairwise test statistics. Its performance will be compared to the LRT proposed by \cite{gail1985testing}. 

The remainder of the paper is structured as follows. In section 2, we provide a review of U-statistics. In section 3, we introduce our proposed U-statistic-based non-parametric test for treatment effect heterogeneity in detail. In section 4, some simulation studies demonstrate the validity of the test and its comparison with the LRT under several different circumstances. In section 5, we apply the proposed method to a randomized study of program effectiveness in labor economics. Additional discussion of this approach can be found in Section 6.\\


\section{ Background: Review of U-Statistics }

U-statistics are a class of statistics widely used to construct non-parametric unbiased estimators of estimable parameters with minimum variance. The asymptotic normality property of U-statistics (under some mild conditions) makes it very popular as a non-parametric testing tool.

We start with a review of one-sample U-statistics \citep{van2000asymptotic}. Let $X_1,\cdots,X_n$ be a random sample from $F(x)$, and assume there is a symmetric function $\phi(x_1,\cdots,x_m)$ $(m \leq n)$ such that $E[\phi(X_1,\cdots,X_m)] = \theta$, where $\theta$ is the parameter of interest. Then the U-statistic for the parameter $\theta $ created by kernel $\phi$ is
\begin{equation}
U(X_1,\cdots,X_n) =\frac{1}{\binom{n}{m}}\sum_{\boldsymbol{\beta }\in B} \phi(X_{\beta_1},\cdots,X_{\beta_m}),
\end{equation}
where $B$ contains all $\binom{n}{m}$ ordered subsets $\boldsymbol{\beta}=(\beta_1,\cdots,\beta_m)$ of $m$ integers chosen without replacement from the set $\{1,\cdots,n\}$ with $1\leq\beta_1<\cdots<\beta_m\leq n$.


The signed rank statistic is an example of a one-sample U-statistic where $\theta = E(I(X_1+X_2>0))$. This  can be used to test whether the location (median) of a symmetric distribution is equal to 0 via testing whether $\theta = \frac{1}{2}$. With the symmetric kernel $\phi(x_1,x_2) = I(x_1+x_2>0)$, the corresponding U-statistic estimator of $\theta$ is 
\begin{equation}
        U = \frac{1}{\binom{n}{2}} \sum_{1\leq i<j \leq n}I(X_i+X_j>0).
\end{equation}

A key property of the one-sample U-statistic is that it has an asymptotic normal distribution. If $E\phi^2(X_1,\cdots,X_m)<\infty$, then
\begin{equation}
\sqrt{n}(U-\theta) \stackrel{D}{\longrightarrow} N(0,\sigma^2),
\end{equation}
where $\sigma^2$ is the asymptotic variance of $\sqrt{n}U$. 

The multi-sample U-statistic is a natural extension of the one-sample U-statistic. 
Let $\{X_{1\alpha}, \alpha = 1,\cdots,n_1\}$, $\cdots$, $\{X_{c\delta}, \delta = 1,\cdots,n_c\}$ be $c$ independent random samples from distribution functions $F_1(x),\cdots,F_c(x)$ respectively, and $\phi(x_{11},\cdots,x_{1m_1};\cdots;x_{c1},\cdots,x_{cm_c})$ be a symmetric function within each set of variables $\{x_{j1},\cdots,x_{jm_j}\}$ $(j = 1,\cdots,c)$ with $E(\phi)$ equal to the parameter of interest $\theta$, where $m_1\leq n_1,\cdots,m_c \leq n_c$. Then the corresponding c-sample U-statistic is 
\begin{equation}
    U = [\prod_{j=1}^c \binom{n_j}{m_j} ]^{-1} \sum_{\boldsymbol{\alpha_1}} \cdots \sum_{\boldsymbol{\alpha_c}} \phi (X_{1\alpha_{1,1}}, \cdots,X_{1\alpha_{1,m_1}}; \cdots;X_{c\alpha_{c,1}}, \cdots,X_{c\alpha_{c,m_c}}) 
\end{equation}
where the summation is over all possible sets of subscripts $\boldsymbol{\alpha_j} = (\alpha_{j,1}, \cdots, \alpha_{j,m_{j}})$ such that $1 \leq \alpha_{j,1}<\cdots<\alpha_{j,m_{j}} \leq n_{j}$ for each of the $c$ samples (i.e, $j=1,\cdots,c$). 

The Mann-Whitney statistic is an example of two-sample U-statistic with $m_1=m_2 =1$. The parameter of interest is $\theta = E(I(X_{11}<X_{21}))$. Under the assumption that the distributions of $X_{1i}$ and $X_{2j}$ are the same up to location shift, $\theta = \frac{1}{2}$ indicates that there is no location shift. The U-statistic with respect to the parameter of interest is 
\begin{equation}
U = \frac{1}{\binom{n_1}{1}\binom{n_2}{1}}\sum_{i=1}^{n_1}\sum_{j = 1}^{n_2}I(X_{1i}<X_{2j}).
\end{equation}

There is also an asymptotic normality property for multi-sample U-statistics, even for a vector of several multi-sample U-statistics defined upon the same sets of mutually independent samples with different kernel functions. \cite{lehmann1963robust} showed that if there are $r$ multi-sample U-statistics $U^{(1)},\cdots,U^{(r)}$, each defined as in (4),  with corresponding kernel functions $\phi^{(1)},\cdots,\phi^{(r)}$ such that $E[\phi^{(k)}]=\theta^{(k)}$ and  $E([\phi^{(k)}]^2)<\infty$ for $k\in\{1,\cdots,r\}$, and if there also exists positive constants $\lambda_j$ $(0<\lambda_j<1)$ such that $\frac{n_j}{N}\rightarrow \lambda_j$ as $N=\sum\limits_{j=1}^c n_j \rightarrow \infty$ for $j\in\{1,\cdots, c\}$, then
\begin{equation}
    \sqrt{N} \left( {\begin{array}{c}
  U^{(1)}-\theta^{(1)} \\
  U^{(2)}-\theta^{(2)}\\
  \vdots\\
  U^{(r)}-\theta^{(r)}
  \end{array} } \right)
  \stackrel{D}{\longrightarrow} N(0, \Sigma),
\end{equation}{}
where $\Sigma$ is the asymptotic covariance matrix of $\sqrt{N}(U^{(1)},U^{(2)},\cdots, U^{(r)})$.

In order to apply (6) to hypothesis testing with regard to the parameter $\theta = (\theta^{(1)}, \cdots, \theta^{(r)})$, we need to identify the form of $\Sigma$. This can be addressed using Hájek projection principle \citep{hajek1968asymptotic} to derive the asymptotic normality property of the U-statistics.

For one c-sample U-statistic of degree $(m_1,\cdots,m_c)$, if $E(\phi^2)<\infty$, the Hájek projection of $U-\theta$ onto the space $\mathcal{V} = \{V|V=\sum\limits_{i=1}^{n_1}f_1(X_{1i})+\dots+\sum\limits_{i=1}^{n_c}f_c(X_{ci})$ where $f_j$ $(j\in\{1,\cdots,c\})$  are some real-valued functions$\}$ is 
\begin{equation}
    \hat{U} = \frac{m_1}{n_1}\sum_{i=1}^{n_1}h_1(X_{1i})+\dots+\frac{m_c}{n_c}\sum_{i=1}^{n_c}h_c(X_{ci}),
\end{equation}{}
where the $h$ functions are defined as
\begin{equation}
    h_j(x) = E[\phi(X_{11},\cdots,X_{1m_1};\cdots;X_{c1},\cdots,X_{cm_c}|X_{j1}=x]-\theta, ~~~ j \in \{1,..,c\}.
\end{equation}{}
Then it can be proved \citep{korolyuk2013theory} that
\begin{equation}
    \sqrt{N}(U-\theta-\hat{U}) \stackrel{P}{\longrightarrow} 0 \text{ as } N\rightarrow \infty.
\end{equation}{}
This shows that $U-\theta$ and $\hat{U}$ have the same asymptotic distribution. By the Central Limit Theorem,
\begin{equation}
    \sqrt{N}\hat{U} \stackrel{d}{\longrightarrow} N(0,\frac{m_1^2}{\lambda_1}Var(h_1(X_{1}))+\cdots+\frac{m_c^2}{\lambda_c}Var(h_c(X_{c}))) \text{ as } N\rightarrow \infty,
\end{equation}{}
provided the variance terms are finite.
Thus we have
\begin{equation}
    \sqrt{N}(U-\theta)\stackrel{D}{\longrightarrow} N(0,\frac{m_1^2}{\lambda_1}Var(h_1(X_{1}))+\cdots+\frac{m_c^2}{\lambda_c}Var(h_c(X_{c}))) \text{ as } N\rightarrow \infty.
\end{equation}{}
With the list of U-statistics $(U^{(1)},\cdots,U^{(r)})$, there is a list of Hájek projection $\hat{U}^{(1)},\cdots,\hat{U}^{(r)}$ corresponding to each of them with 
\begin{equation}
    \hat{U}^{(k)} = \frac{m_1^{(k)}}{n_1}\sum_{i=1}^{n_1}h_1^{(k)}(X_{1i})+\dots+\frac{m_c^{(k)}}{n_c}\sum_{i=1}^{n_c}h_c^{(k)}(X_{ci}) \text{ for } k\in\{1,\cdots,r\},
\end{equation}{}
where $h_j^{(k)}(x) = E[\phi^{(k)}(X_{11},\cdots,X_{1n_1};\cdots;X_{c1},\cdots,X_{cn_c}|X_{j1}=x]-\theta^{(k)}$ for $j\in\{1,\cdots,c\}$. By the multidimensional Central Limit Theorem, we know
\begin{equation}
    \sqrt{N}
    \left( {\begin{array}{c}
  \hat{U}^{(1)} \\
  \hat{U}^{(2)}\\
  \vdots\\
  \hat{U}^{(r)}
  \end{array} } \right)
  \stackrel{D}{\longrightarrow} N(0, \frac{1}{\lambda_1}\Sigma_1+\cdots+\frac{1}{\lambda_c}\Sigma_c)
\end{equation}{}
where $\Sigma_j = Cov[m_j^{(1)}h_j^{(1)}(X_j),\cdots,m_j^{(r)}h_j^{(r)}(X_j)]$ for $j\in\{1,\cdots,c\}$. Since

\begin{equation}
\sqrt{N}
    \left( {\begin{array}{c}
  U^{(1)}-\theta^{(1)}-\hat{U}^{(1)} \\
  U^{(2)}-\theta^{(2)}-\hat{U}^{(2)}\\
  \vdots\\
   U^{(r)}-\theta^{(r)}-\hat{U}^{(r)}
  \end{array} } \right)
  \stackrel{P}{\longrightarrow} 0 \text{ as } N\longrightarrow\infty,
\end{equation}{}
we have 
\begin{equation}
    \sqrt{N}
    \left( {\begin{array}{c}
  U^{(1)}-\theta^{(1)} \\
  U^{(2)}-\theta^{(2)}\\
  \vdots\\
   U^{(r)}-\theta^{(r)}
  \end{array} } \right)
  \stackrel{D}{\longrightarrow} N(0, \frac{1}{\lambda_1}\Sigma_1+\cdots+\frac{1}{\lambda_c}\Sigma_c).
\end{equation}{}

\section{Testing for Treatment Effect Heterogeneity}
Suppose we are focused on a study population comprised of $S$ strata. For each stratum $s, s\in\{1,\cdots,S\}$, let $Y_s^t$ denote the outcomes of subjects in the  treatment group where $Y_s^t = \{Y_{si}^t,i=1,\cdots,n_s^t\}$, and $Y_s^c$ denotes the outcomes in the control group where $Y_s^c=\{Y_{si}^c,i=1,\cdots,n_s^c\}$. Define $N_s=n_s^t+n_s^c$ as the total sample size in strata $s$, and $N=\sum_{s=1}^S N_s$ as the overall sample size across all strata. We develop a non-parametric U-statistic-based test (U test) for the null hypothesis of no treatment effect heterogeneity against the alternative hypothesis that not all treatment effects are equal. In the derivation and in our studies, we focus on an assumed additive treatment effect. Alternative methods of the treatment effect can be considered, they would require alternative choices for the U-statistic kernel functions.

The technique to be discussed here relies on two assumptions: (1) $Y_1^t,\cdots,Y_S^t,Y_1^c,\cdots,Y_S^c$ are mutually independent; (2)There exist constants $\lambda_s^{\omega} \in (0, 1)$ such that $\frac{n_s^{\omega}}{N}\rightarrow \lambda_s^\omega$  for all $s\in\{1,\cdots,S\}$ and $\omega \in \{t, c\}$.


\subsection{Comparing Treatment Effects between the First Two Strata}
We start by constructing a U-statistic comparing the treatment effects of the first two strata. Note that under the null hypothesis of equal additive treatment effect, $E[I(Y_{1}^t-Y_{1}^c<Y_{2}^t-Y_{1}^c)+ \frac{1}{2}I(Y_{1}^t-Y_{1}^c = Y_{2}^t-Y_{2}^c)]$ will be equal to $\frac{1}{2}$. The second term is included to account for the possibility of ties for discrete distributions. Then the 4-sample U-statistic based on kernel function $\phi^{(1,2)}(y_1^t;y_1^c;y_2^t;y_2^c) = I(y_1^t-y_1^c<y_2^t-y_2^c)+\frac{1}{2}I(y_1^t-y_1^c = y_2^t-y_2^c)$ is 
\begin{equation}
     U^{(1,2)} = \frac{1}{n_{1}^tn_{1}^cn_{2}^tn_{2}^c}\sum_{i=1}^{n_{1}^t}\sum_{j=1}^{n_{1}^c}\sum_{k=1}^{n_{2}^t}\sum_{l=1}^{n_{2}^c}I(Y_{1i}^t-Y_{1j}^c<Y_{2k}^t-Y_{2l}^c) +\frac{1}{2}I(Y_{1i}^t-Y_{1j}^c = Y_{2k}^t-Y_{2l}^c).
\end{equation}

Denoting $\theta^{(1,2)} = E(U^{(1,2)})$ and using the background results about multi-sample U-statistics with $r=1$, with the assumption that $\frac{n_s^\omega}{N}\rightarrow \lambda_s^\omega (0<\lambda_s^\omega<1)$ as $N\rightarrow \infty$ and the fact that $E[(\phi^{(1,2)})^2]\leq 1$
, we have
\begin{equation}
       \sqrt{N}(U^{(1,2)}-\theta^{(1,2)}) \stackrel{D}{\longrightarrow} N(0,\sigma^2_{1,2}),
\end{equation}
where 
\begin{align*}
        &\sigma^2_{1,2} = \frac{1}{\lambda_1^t}Var(h_1^{t,(1,2)}(Y_1^t))+
        \frac{1}{\lambda_1^c}Var(h_1^{c,(1,2)}(Y_1^c))+\frac{1}{\lambda_2^t}Var(h_2^{t,(1,2)}(Y_2^t))+
        \frac{1}{\lambda_2^c}Var(h_2^{c,(1,2)}(Y_2^c)) \in (0,\infty),\\
        &h_s^{\omega,(1,2)}(x) = E[\phi^{(1,2)}(Y_{1}^t;Y_{1}^c;Y_{2}^t;Y_{2}^c)|Y_s^\omega = x]-\theta^{(1,2)} \text{ for } s \in\{1,2\}, \omega\in\{t,c\}.
\end{align*}{}

To apply this method, we first estimate $h_s^{\omega,(1,2)}(x)$ ($s\in\{1, 2\}$ and $\omega \in \{t, c\}$), an expectation, by the method of moments. For instance, $h_1^{t,(1,2)}(x)$ is estimated by the sample mean $\hat{h}_1^{t,(1,2)}(x) = \frac{1}{n_{1}^cn_{2}^tn_{2}^c}\sum\limits_{j=1}^{n_{1}^c}\sum\limits_{k=1}^{n_{2}^t}\sum\limits_{l=1}^{n_{2}^c}I(x-Y_{1j}^c<Y_{2k}^t-Y_{2l}^c)$. Note that this calculation is repeated with each data value $Y_{1i}^t$ $(i = 1,\cdots,n_1^t)$ taking the place of $x$. Likewise for other $h$ terms. Then we estimate $Var(h_s^{\omega,(1,2)}(Y_s^\omega))$ by the sample variance of $\hat{h}^{\omega,(1,2)}_s(Y_{s}^\omega)$ as $\frac{1}{n_s^\omega-1}\sum\limits_{i=1}^{n_s^\omega}[\hat{h}_s^{\omega,(1,2)}(Y^{\omega}_{si}) - \frac{1}{n_s^\omega} \sum\limits_{j = 1}^{n_s^\omega}\hat{h}_s^{\omega,(1,2)}(Y_{sj}^\omega)]^2$, for $s\in\{1,2\}$ and $\omega \in \{t,c\}$, and take the weighted sum of them to approximate $\sigma^2_{1,2}$.

\subsection{Testing Treatment Effect Heterogeneity across Multiple Strata}

With S strata, we can construct a test statistic like (16) for any pair of strata. Denote $U^{(p,q)}(p<q)$ as the U-statistic comparing strata $p$ and $q$ with kernel $\phi^{(p,q)}(y_p^t;y_p^c;y_q^t;y_q^c)=I(y_p^t-y_p^c<y_q^t-y_q^c) + \frac{1}{2}I(y_p^t-y_p^c=y_q^t-y_q^c)$ and expectation $\theta^{(p,q)}$. By applying the Hájek projection principle to a vector of multi-sample U-statistics as in Section 2, we have

\begin{equation}
\sqrt{N}
    \left( {\begin{array}{c}
  U^{(1,2)}-\theta^{(1,2)} \\
  U^{(1,3)}-\theta^{(1,3)}\\
  \vdots\\
   U^{(S-1,S)}-\theta^{(S-1,S)}
  \end{array} } \right)
  \stackrel{D}{\longrightarrow} N(0, \Sigma)  
\end{equation}{}
where $\Sigma = \frac{1}{\lambda_1^t}\Sigma_1^t+\frac{1}{\lambda_1^c}\Sigma_1^c+\dots+\frac{1}{\lambda_S^t}\Sigma_S^t+\frac{1}{\lambda_S^c}\Sigma_S^t$ and\\ $\Sigma_s^\omega = Cov(\Tilde{h}_s^{\omega,(1,2)}(Y_s^\omega),\cdots,\Tilde{h}_s^{\omega,(1,S)}(Y_s^\omega),\Tilde{h}_s^{\omega,(2,3)}(Y_s^\omega),\cdots,\Tilde{h}_s^{\omega,(S-1,S)}(Y_s^\omega))$ for all $s\in\{1,\cdots,S\}$ and $\omega \in \{t,c\}$. 
Here 
\begin{equation*}
  \Tilde{h}_s^{\omega,(p,q)}(x) = 
    \begin{cases}
     h_s^{\omega,(p,q)}(x)=  E[\phi^{(p,q)}(Y_p^t;Y_p^c;Y_q^t;Y_q^c)|Y_s^\omega = x]-\theta^{(p,q)}  & \text{ if } s=p \text{ or } s=q, \\
     0 & o.w. \\
    \end{cases}
\end{equation*}
for $\{(p,q)|1\leq p<q\leq S\}$. Under $H_0:\tau_1 = \cdots = \tau_S$, all $\theta$'s are equal to $\frac{1}{2}$.

Estimation of $\Sigma$ is carried out using a similar approach as described for estimation of $\sigma_{1,2}^2$ in Section 3.1. We first construct an empirical estimate for each  $h$ function (as in the paragraph below (17)) and then use the sample covariance matrix of\\ $[\Tilde{h}_s^{\omega,(1,2)}(Y_s^\omega),\cdots,\Tilde{h}_s^{\omega,(1,S)}(Y_s^\omega),\Tilde{h}_s^{\omega,(2,3)}(Y_s^\omega),\cdots,\Tilde{h}_s^{\omega,(S-1,S)}(Y_s^\omega)]$ with $h's$ replaced by their corresponding estimates to get $\hat{\Sigma}_s^\omega$. Then $\hat{\Sigma}$, the estimate of $\Sigma$, is the sum of the $\hat{\Sigma}_s^\omega$ over $s\in\{1,\cdots,S\}$ and $\omega \in\{t,c\}$ with each term weighted by $\frac{1}{\lambda_s^{\omega}}$.

The vector of pairwise test statistics $U = (U^{(1,2)}, U^{(1,3)},\cdots, U^{(S-1, S)})^T$ can be combined into a single overall test statistic using any function of $U$. Here we focus on $U_h = N\cdot \sum\limits_{1\leq p<q\leq S}(U^{(p,q)}-\frac{1}{2})^2$. The asymptotic distribution of $U_h$ is not available in analytic form, but a simulation approach can be used to assess $U_h$. A large number of independent samples of $\sqrt{N}(U-\frac{1}{2})$ are generated from the null $N(0,\hat{\Sigma})$ distribution, and $U_h$ is computed for each sample to generate the empirical null distribution of $U_h$. For $\alpha$ level test, we reject $H_0$ when $U_h$ is greater than or equal to the $100(1-\alpha)$ percentile of the empirical null distribution. 
Note that other test statistics are also possible, e.g., $\sqrt{N} \cdot \max\limits_{1\leq p<q\leq S}|U^{(p,q)}-\frac{1}{2}|$, and simulation is always an option for deriving the reference distribution.
In our simulation study, we use the statistic $U_h$ because it proved reliable.

Another test statistic that might seem natural is $T = N(U-\frac{1}{2}\mathbf{1})\hat{\Sigma}^-(U-\frac{1}{2}\mathbf{1})^T$ whose reference distribution is $\chi_k^2$, where $k$ is the rank of $\hat{\Sigma}$. $\hat{\Sigma}^- = \sum\limits_{i = 1}^k \frac{1}{\alpha_i} q_iq_i^t$ is a generalized inverse of $\hat{\Sigma}$, where $ \{\alpha_i, i = 1,\cdots,k\}$ are the non-zero eigenvalues of $\hat{\Sigma}$ and $ \{q_i, i = 1,\cdots,k\}$ are the corresponding orthogonal eigenvectors. However, as $U$ is a vector of all of the pairwise U-statistics, the determinant of its covariance matrix can be very close to 0. Then $\hat{\Sigma}$ can have an eigenvalue $\alpha$ very close to 0, for which a tiny rounding error would have a large impact on $\hat{\Sigma}^-$ and thus on $T$. So even though the reference distribution of the test statistic $T$ has known distribution, we prefer using $U_h$.

\subsection{Three Particular Cases to Apply}
The approach described above does not make any specific assumption about the shapes of the distributions of $Y_1^t,\cdots, Y_S^t, Y_1^c,\cdots,Y_S^c$ or the relationship among them other than independence. Our U-statistic is testing whether the probability $P(Y_p^t-Y_p^c<Y_q^t-Y_q^c)+\frac{1}{2}P(Y_p^t-Y_p^c=Y_q^t-Y_q^c)~~(p\neq q)$ is equal to one half for all $p$, $q$. This does not however provide much insight into the feature of outcomes that is being tested. When there is no further assumptions of the outcome distributions at all, confusion may arise. We describe three cases here, with a set of assumptions for each of them, under which the interpretation of the test is clear. Many real-life problems may fit into those cases.

\textit{Case A}: We assume all outcomes in different strata and different treatment groups follow a common distribution $F$ up to a location shift, which is comprised of the additive treatment effects $\tau_s$ within each stratum and assumed additive stratum effects $\Delta_s$ for $s\in\{1,\cdots,S\}$. In this case, $P(Y_p^t-Y_p^c<Y_q^t-Y_q^c)+\frac{1}{2}P(Y_p^t-Y_p^c=Y_q^t-Y_q^c)$ is equal to a half if and only if the within-stratum location shifts ($\tau_p$ and $\tau_q$) are the same. It can then be easily shown that our test is consistent with identifying whether $\tau_s$ is the same across all strata. Since the shift can be considered as the difference in means or difference in any percentiles between two different treatment groups, the test is consistent with respect to the alternative hypothesis that the difference of means (or percentiles) are unequal for at least one pair of strata. This case is equivalent to testing the interaction in a two-factor factorial design when one factor has two levels; in our case, one factor is stratum and the other is treatment which takes two levels. The ANOVA F-test can be used to address this scenario, and non-parametric tests have been proposed as an alternative \citep{de2017mann, patel1973nonparametric}. The \cite{de2017mann} approach is similar to our approach in that they use the same basic U-statistic that we do. They use a different summary statistic to aggregate the pairwise comparison. A limitation is that their approach applies only to this case (\textit{Case A}) and not to \textit{Case B} or \textit{Case C} that are described next.

\textit{Case B}: We assume that all outcomes in the treatment groups follow a common distribution $F^t$ up to a stratum-specific location shift $\Delta_s^t$ and all outcomes in the control groups follow a possibly different common distribution $F^c$ up to a stratum-specific location shift $\Delta_s^c$ for all $s\in\{1,\cdots,S\}$. Then the within-stratum treatment effect can be depicted as the difference of the location shift $\Delta_s^t-\Delta_s^c$ plus the difference of $F^t$ and $F^c$. No matter what metric is used to describe the difference of $F^t$ and $F^c$, it is consistent across all strata. So in this case, $P(Y_p^t-Y_p^c<Y_q^t-Y_q^c)+\frac{1}{2}P(Y_p^t-Y_p^c=Y_q^t-Y_q^c)$ is $\frac{1}{2}$ if and only if $\Delta_p^t-\Delta_p^c = \Delta_q^t-\Delta_q^c$. So our test is identifying whether the difference of location shifts $\Delta_s^t-\Delta_s^c$ are consistent across strata, which can also be considered as the consistency of the difference of means or percentiles between different treatment groups. We can easily prove the test is consistent with the alternative hypothesis that the difference of means (or percentiles) between the treatment and control groups are not all identical across strata. 

\textit{Case C}: We assume that the outcomes in the same strata $s \in \{1,\cdots,S\}$ follow the same distribution $F_s$ up to a stratum-specific location shift $\tau_s$, the additive treatment effect. Similarly, in this case, our method is a consistent test 
with respect to the alternative hypothesis that 
there are  at least two strata such that the difference of means (or percentiles) between the treatment and control groups are different.


\textit{Case A} can be applied in practice whenever we think the factor we use to stratify subjects, as well as the treatment, can only shift the location of the outcome distributions but cannot change the shape. This is a common assumption. \textit{Case B} can be applied when the treatment changes the shape of the outcome distributions, whereas the factor used to stratify subjects would only shift the location. Suppose we are studying whether the effect of a welfare reform policy on household income differs across several different geographic regions. Usually a welfare reform plan has different magnitude of impact on families who are on different economic levels. Thus we may expect the distribution of income after the reform to differ from the control condition. If the shape of income distribution is similar across geographic regions or just differs by scale among different regions, then this scenario could fit in our \textit{Case B}. \textit{Case C} can be applied when the treatment effects are constant within each stratum, but the distribution of outcomes differs across strata. Back to the welfare reform example. If we stratify subjects according to their economic levels, then income distributions would be expected to vary among different strata, but it is possible that treatment effects could be constant.

\section{Simulation Study}
We demonstrate the U-statistic test of treatment effect heterogeneity via a simulation study. There are some computational challenges that are addressed first, then the simulation study is described and results are provided. Simulations compare the U test to the LRT in a range of scenarios, some of which match the assumptions in the parametric test and others do not. Additionally, we show the relationship among power, sample size and effect size in two scenarios.

\subsection{Computational Issues}
When the sample sizes are large, the computation of the U-statistic is computationally expensive. Let's take $U^{(1,2)}$ as an example. We need to compute the average over all the combinations of $I(Y_{1i}^t-Y_{1j}^c<Y_{2k}^t-Y_{2l}^c) +\frac{1}{2}I(Y_{1i}^t-Y_{1j}^c = Y_{2k}^t-Y_{2l}^c)$, denoted by $\phi^{(1,2)}(i,j,k,l)$, which includes $n_1^t \times n_1^c \times n_2^t \times n_2^c$ terms. As this computation can be done in parallel for different $(i,j,k,l)$, it should not be a big problem when applying the method for a single data set. In simulations, we need to generate thousands of data sets and compute U-statistics for each of them. So in the simulation study, 
instead of computing the average through exhaustive enumeration, we generate approximate U-statistics by randomly selecting some of the combinations with replacement and use this average to approximate the U-statistic. We randomly selected $M = 10^3N$ samples with replacement from each treatment subgroup as $\{y_{1i}^t,y_{1i}^c,y_{2i}^t,y_{2i}^c, i = 1,\cdots,M\}$ to approximate $U^{(1,2)}$, here $N$ is the total sample size of the two strata. The sampling size $M$ was determined by considering a range of scenarios and estimating the variance of the test statistic $U_h = N\cdot \sum\limits_{1\leq i<j\leq S}(U^{(i,j)}-\frac{1}{2})^2$ constructed by approximate U-statistics within each scenario via simulation.  The variance increases as $N$ increases. In the simulation, as $N$ ranged from 60 to 3000, the maximum estimated variances for this choice of $M$ ranged from 0.0058 to 0.104, which was judged to provide sufficient precision. One requirement of this sampling is that all subjects have to be selected at least once, 
because we also used the sampled indicators to estimate $h_s^{\omega,(1,2)}(Y_{si}^{\omega})$ $(s\in\{1,2\}, \omega \in \{t,c\}, i\in\{1,\cdots,n_s^{\omega} \})$. 
For instance, the estimate of $h_1^{t,(1,2)}(Y_{11}^t)$ is computed as the average of all selected $\phi^{(1,2)}(i,j,k,l)$ with $Y_{1i}^t$ equal to $Y_{11}^t$. 
Though this requirement is not a challenge due to the large sampling size $M$, in the rare events that it occurs we would need to redo the sampling procedure. As for the empirical reference distribution of the test statistic $U_h$, we generated $10^5$ random samples $r_i = (r_{i}^{(1,2)},\cdots,r_{i}^{(S-1,S)})$ $(i = 1,\cdots,10^5)$ from
the multivariate normal distribution $N(0,\hat{\Sigma})$ for each simulation, and got the distribution of $||r_i||^2$ as the empirical reference distribution under $H_0$. The empirical p-value is the percentage of generated $||r_i||^2$ greater than $U_h$. Fixing the type I error as $\alpha = 0.05$, we reject $H_0$ when the p-value is smaller than $\alpha$. We determined the sample size $10^5$ by considering a range of scenarios and estimating the variance associated with a simulation-based 95th percentile. The sample size of $10^5$ in the various cases makes the variance less than 0.01.

\subsection{Review of the Likelihood Ratio Test for Treatment Effect Heterogeneity}
The likelihood ratio test for treatment effect heterogeneity was developed by \cite{gail1985testing}. 
Let $\tau_s$ denote the treatment effect in subgroup $s$ $(s\in \{1,\cdots,S\})$. The test assesses the null hypotheses $H_0: \tau_1=\cdots=\tau_S$ versus the alternative that at least two of the subgroup treatment effects are unequal. Under the assumption that $\Hat{\tau}_s (s\in \{1,\cdots,S\})$ follows a normal distribution with
\begin{equation}
    \hat{\tau_s} \stackrel{indep}{\sim} N(\tau_s, \sigma_s^2), s\in\{1,\cdots,S\},
\end{equation}
we have heterogeneity test statistic
\begin{align}
    H &= \sum_{s=1}^S(\hat{\tau}_s-\bar{\hat{\tau}})^2/s_s^2 \stackrel{H_0}{\sim} \chi_{S-1}^2\\
    \text{where } &\bar{\hat{\tau}} = (\sum_{s=1}^S \hat{\tau}_s/s_s^2)/(\sum_{s=1}^S 1/s_s^2), \text{and $s_s^2$ is a consistent estimator of  $\sigma_s^2$.}\nonumber
\end{align}
With fixed type I error $\alpha$, we reject the null hypotheses $H_0$ when the test statistic $H$ is greater than or equal to the $100(1-\alpha)th$ percentile of $\chi_{S-1}^2$. 

For additive treatment effects, the treatment effect estimates $\hat{\tau}_s$ can be the difference between the sample means of the two treatment groups within strata $s$. When the subgroup sample sizes are large, according to the Central Limit Theorem, $\hat{\tau}_s$ will approximate to a normal distribution. However, when the distributions of the outcomes differ from normality and the validity of the test relies on large sample sizes, the power of the test will be impacted as with other parametric tests \citep{lehmann2004elements}.

\subsection{Simulation Study Design}

Assuming we have three strata, we generated $ n_s^\omega $ random samples from treatment subgroup $\omega$ within strata $s$ from a distribution $F_s^\omega$ $(s \in {1,2,3}; \omega \in \{t,c\})$. The choices of $n_s^\omega$ and $F_s^\omega$ are described below. The hypothesis of no heterogeneity was tested via our non-parametric U test and the LRT reviewed in the previous section. For each simulation scenario (choices of $n_s^\omega$ and $F_s^\omega$), we generate $L = 2000$ data sets and carry out the tests on each. This yields rejection rates and the empirical distribution of p-values. 

We developed 17 different scenarios for the choice of the distributions $F_s^\omega$.  These scenarios are listed in Table 1. They are organized according to the three application cases outlined in Section 3.3. For each scenario, the true treatment effects were varied to provide the null and alternative instances. The upper half of Table 1 lists the null cases and the lower half lists the alternative cases.

For \textit{Case A}, outcomes in the three strata and two treatment groups follow a common distribution $F$ up to location shifts. The first scenario ($A1$) is that $F = N(0,1)$. For the next two scenarios ($A2$ and $A3$), $F$ are still symmetric distributions, but with tails lighter ($A2$) or heavier ($A3$) than normal distribution. Next we consider three skewed distributions $\chi_1^2$, $Exp(1)$ and $\chi_4^2$ (labeled as $A4-A6$) with their skewnesses decreasing in that order. The support of these distributions are all positive. We are generally more interested in comparing the scales of the treatment and control groups instead of location shifts in these scenarios. So we suppose there are constants $c_s^\omega$ such that $\frac{Y_s^\omega}{c_s^\omega} \sim F$ $(s \in {1,2,3}; \omega \in \{t,c\})$. Here we use the logarithm of those outcomes in our test statistic $U_h$ instead of using the original outcomes directly. Now the problem of testing the consistency of ratio of scales is changed into a problem of testing the consistency of the location shift $log(c_s^t) - log(c_s^c)$ $(s = 1,\cdots,S)$. The final \textit{Case A} example is a bimodal distribution $0.5N(-5,1)+0.5N(5,1)$ (labeled as $A7$). 

In \textit{Case B}, all outcomes in the treatment groups follow a common distribution $F^t$ up to a location shift and all outcomes in the control group follow a different common distribution $F^c$ up to a location shift. To create examples here, we choose two of the distributions used in \textit{Case A} whose supports are the whole real line ($A1-A3$, $A7$) and randomly assign them to the treatment and control group. We try all $\binom{4}{2}$ combinations, they are labeled as $B1-B6$.

In \textit{Case C} where all outcomes in the same stratum follow a common distribution $F_s$ ($s = 1,\cdots,S$) up to a location shift, we select three distribution from \textit{Case A} with support on the whole real line and randomly assign them to the three strata. So we have $\binom{4}{3}$ combinations and they are labeled as scenarios $C1-C4$.

For each scenario described above, we vary the true treatment effects to get the null and alternative cases, and also consider a range of different sample sizes. In each scenario, there are stratum-specific location shifts ($\Delta_1 = 0, \Delta_2 = 1, \Delta_3 = 2$ for scenarios in \textit{Case A} and \textit{Case B}). For all null cases, the stratum treatment effects are the same across the three strata. For alternative cases, the treatment effects $( \tau_1, \tau_2, \tau_3)$ form an arithmetic series with $\tau_2 = \tau_1 + \Gamma$ and $\tau_3 = \tau_1 + 2\Gamma$, where $\Gamma >0$.
Simulations were carried out for a range of values of $\Gamma$. If $\Gamma$ is too large, then both tests always reject the null hypothesis for almost any sample size. Results are presented for a representative choice of $\Gamma$ where this does not occur.
All simulation scenarios were investigated with six assumptions regarding sample sizes. First, all subgroup sample sizes are the same $(n_s^{\omega} = n)$ with $n$ equal to 10, 50, 100 and 500. Second, the sample sizes of treatment and control within each stratum are the same, but sample size varies across strata. We tried $(n_1^\omega, n_2^\omega, n_3^\omega)$ $(\omega \in\{t, c\})$ equal to $(50,100,150)$ and  $(150,100,50)$. The former corresponds to the case that the strata with higher effect sizes have larger sample sizes. The latter is the opposite case.

\subsection{Simulation Study Results}

The rejection rates of both the U test and the LRT assuming $\alpha = 0.05$ for all null cases of the different scenarios are provided in Table 2.  The first column of Table 2 shows the labels of all scenarios. Each of the remaining columns corresponds to one sample size setting for all scenarios. As expected all rejection rates are close to the 0.05 level. Since empirical type I errors were computed by generating data for $L = 2000$ times, the standard error for each is approximately 0.005. The table shows that when $n_s^\omega = 10$ $(s\in\{1,2,3\};\omega \in \{t,w\})$, the type I errors of both tests are a bit too high in all scenarios except for $A4$ and $C4$. For scenario $C3$, when $(n_1, n_2, n_3)$ is equal to $(50, 50, 50)$ or $(150,100,50)$, type I errors of both tests are a bit too high. In all other settings, the type I errors are well controlled for both tests.

We then compare the power of the two tests by comparing their rejection rates for the alternative cases in all sample size settings. Figure 1, 2 and 3 show the results for scenarios in \textit{Case A}, \textit{Case B} and \textit{Case C} separately. Each figure is comprised of two subfigures (a) and (b). Subfigure (a) shows rejection rates for all cases where sample sizes $n_s^\omega$  $(s\in\{1,2,3\})$ are equal. Subfigure (b) focuses on the three cases whose stratum-specific sample sizes can vary. Within each subfigure, there is a set of panes, each of which corresponds to a scenario. Within each pane, the vertical axis indicates the rejection rate and the horizon axis indicates the sample size setting. For each type of test, we plot a point showing the empirical rejection rate and a line showing the corresponding 95\% confidence interval. The red ones are for our proposed U test, and the blue ones are for the LRT.  

For \textit{Case A}, Figure 1(a) shows that when stratum-specific sample sizes are equal, the powers of both tests increase as $n$ increases. When the common distribution $F$ is normal ($A1$) or $F$ is symmetric with tails lighter than normal ($A2$), the power of the LRT is a bit higher than the U test. When $F$ is symmetric with tails heavier than normal ($A3$), or $F$  is skewed ($A4$ - $A6$) or bimodal ($A7$), the U test is more powerful than the LRT. Also as $F$ departs more from the Gaussian distribution, the advantage of the U test over the LRT is more substantial. Figure 1(b) shows the cases when subgroup sample sizes average 100 but vary across the strata. Compared to the cases with equal sample sizes, the power of both the U test and the LRT drop, and the U test power drops a bit more than the LRT. 

Figure 2 shows the rejection rates of the U test and the LRT for \textit{Case B}. The results of cases with equal sample sizes across strata are in Figure 2(a). As the sample size increases, the powers of both tests increase in all scenarios. When the distributions in both treatment groups are close to normal ($B1$), the LRT is more powerful, otherwise the U test is more powerful. When one of the distributions is very far from normal ($B3$, $B5$ and $B6$), the advantage of the U test over the LRT is large. Next we compare the cases with average subgroup sample sizes all equal to 100 but where sample sizes can vary across strata (Figure 2(b)). As with \textit{Case A}, the results with different stratum-specific sample sizes indicate less power than the case with equal stratum-specific sample sizes for both the U test and the LRT.

The empirical power of the two tests for \textit{Case C} are displayed in Figure 3. When the stratum-specific sample sizes are equal (Figure 3(a)), the power of both tests increase as sample size increases. In all scenarios, the U test outperforms the LRT, and when there is a bimodal distribution ($C2$, $C3$, $C4$), the advantage of the U test is substantial. Figure 3(b) shows the effect of varying sample sizes across strata. When the distributions of the data in the three strata have similar variances and none of them are too far from normal ($C1$), the comparison result is similar to that in \textit{Case A} and \textit{Case B}. The setting with consistent stratum-specific sample sizes has largest power for both the U test and the LRT, and the difference of powers between balanced sample size setting and unbalanced sample size setting is larger for the U test than the LRT. When one of the strata follows a distribution that is mixture normal ($0.5N(-5,1) + 0.5N(5,1)$) which has a lot larger variance than the other two distributions and also departs more from the normal, the performances of the two tests are very different. For the U test, if the stratum with large-variance distribution has the smallest sample size, it is least powerful. For the LRT, when the stratum with distribution very far from normal has the largest sample size, it is least powerful.

\begin{table}[H]
    \centering
    \scalebox{0.46}{
    \begin{tabular}{|c|c|c|c|c|c|c|}
     \hline
  Scenario&$F_1^t$ & $F_1^c$ &$F_2^t$ & $F_1^c$ & $F_3^t$ &$F_3^c$ \\
      \hline
       Null Cases \\
       \hline
        $A1$ & $N(0,1)$&$N(0,1)-1$&$N(0,1)+1$&$N(0,1)$&$N(0,1)+2$&$N(0,1)+1$\\
         $A2$ & $U(-2,2)$&$U(-2,2)-1$&$U(-2,2)+1$&$U(-2,2)$&$U(-2,2)+2$&$U(-2,2)+1$\\
         $A3$ & $t_4$&$t_4-1$&$t_4+1$&$t_4$&$t_4+2$&$t_4+1$\\
       $A4$& $e^1\cdot\chi^2_1$&$\chi^2_1$&$e^2\cdot\chi^2_1$&$e^1\cdot\chi^2_1$&$e^3\cdot\chi^2_1$&$e^2\cdot\chi^2_1$\\
         $A5$ & $e^1\cdot Exp(1)$&$Exp(1)$&$e^2\cdot Exp(1)$&$e^1\cdot Exp(1)$&$e^3\cdot Exp(1)$&$e^2\cdot Exp(1)$\\
         $A6$& $e^1\cdot\chi^2_4$&$\chi^2_4$&$e^2\cdot\chi^2_4$&$e^1\cdot\chi^2_4$&$e^3\cdot6\chi^2_4$&$e^2\cdot\chi^2_4$\\
          $A7$& $0.5N(-5,1)+0.5N(5,1)$&$0.5N(-5,1)+0.5N(5,1)-1$&$0.5N(-5,1)+0.5N(5,1)+1$&$0.5N(-5,1)+0.5N(5,1)$&$0.5N(-5,1)+0.5N(5,1)+2$&$0.5N(-5,1)+0.5N(5,1)+1$\\
         \hline
         $B1$ & $N(0,1)$ & $U(-2,2)$ &  $N(0,1)+1$ & $U(-2,2)+1$ & $N(0,1)+2$ & $U(-2,2)+2$\\
         $B2$ & $N(0,1)$ & $t_4$ &  $N(0,1)+1$ & $t_4+1$ & $N(0,1)+2$ & $t_4+2$\\
         $B3$ & $N(0,1)$ & $0.5N(-5,1)+0.5N(5,1)$ &  $N(0,1)+1$ & $0.5N(-5,1)+0.5N(5,1)+1$ & $N(0,1)+2$ & $0.5N(-5,1)+0.5N(5,1)+2$\\
         $B4$ & $U(-2,2)$ & $t_4$ &  $U(-2,2)+1$ & $t_4+1$ & $U(-2,2)+2$ & $t_4+2$\\
         $B5$ & $U(-2,2)$ & $0.5N(-5,1)+0.5N(5,1)$ &  $U(-2,2)+1$ & $0.5N(-5,1)+0.5N(5,1)+1$ & $U(-2,2)+2$ & $0.5N(-5,1)+0.5N(5,1)+2$\\
         $B6$ & $t_4$ & $0.5N(-5,1)+0.5N(5,1)$ &  $t_4+1$ & $0.5N(-5,1)+0.5N(5,1)+1$ & $t_4+2$ & $0.5N(-5,1)+0.5N(5,1)+2$\\
         \hline
         $C1$ & $N(0,1)$ & $N(0,1)-1$ &  $U(-2,2)$ & $U(-2,2)-1$ & $t_4$ & $t_4-1$ \\
         $C2$ & $N(0,1)$ & $N(0,1)-1$ &  $U(-2,2)$ & $U(-2,2)-1$ & $0.5N(-5,1)+0.5N(5,1)$ & $0.5N(-5,1)+0.5N(5,1)-1$ \\
         $C3$ & $N(0,1)$ & $N(0,1)-1$ &  $t_4$ & $t_4-1$ & $0.5N(-5,1)+0.5N(5,1)$ & $0.5N(-5,1)+0.5N(5,1)-1$ \\
         $C4$ & $U(-2,2)$ & $U(-2,2)-1$ &  $t_4$ & $t_4-1$ & $0.5N(-5,1)+0.5N(5,1)$ & $0.5N(-5,1)+0.5N(5,1)-1$ \\
         \hline
        Alternative Cases\\
        \hline
        $A1$ & N(0,1)&N(0,1)-1&N(0,1)+1&N(0,1)+1-1.25&N(0,1)+2&N(0,1)+2-1.5\\
       $A2$ & $U(-2,2)$&$U(-2,2)-1$&$U(-2,2)+1$&$U(-2,2)+1-1.1$&$U(-2,2)+2$&$U(-2,2)+2-1.2$\\
       $A3$ & $t_4$&$t_4-1$&$t_4+1$&$t_4+1-1.25$&$t_4+2$&$t_4+2-1.5$\\
         $A4$& $e^1\cdot\chi^2_1$&$\chi^2_1$&$e^{2.5}\cdot\chi^2_1$&$e^1\cdot\chi^2_1$&$e^4\cdot9\chi^2_1$&$e^2\cdot\chi^2_1$\\
        $A5$ & $e^1\cdot Exp(1)$&$Exp(1)$&$e^{2.25}\cdot Exp(1)$&$e^1\cdot Exp(1)$&$e^{3.5}\cdot Exp(1)$&$e^2\cdot Exp(1)$\\
        $A6$& $e^1\cdot\chi^2_4$&$\chi^2_4$&$e^{2.25}\cdot\chi^2_4$&$e^1\cdot\chi^2_4$&$e^{3.5}\cdot\chi^2_4$&$e^2\cdot\chi^2_4$\\
        $A7$& $0.5N(-5,1)+0.5N(5,1)$&$0.5N(-5,1)+0.5N(5,1)-1$&$0.5N(-5,1)+0.5N(5,1)+1$&$0.5N(-5,1)+0.5N(5,1)+1-2$&$0.5N(-5,1)+0.5N(5,1)+2$&$0.5N(-5,1)+0.5N(5,1)+2-3$\\
        \hline
         $B1$ & $N(0,1)$ & $U(-2,2)$ &  $N(0,1)+1$ & $U(-2,2)+1-0.25$ & $N(0,1)+2$ & $U(-2,2)+2-0.5$ \\
         $B2$ & $N(0,1)$ & $t_4$ &  $N(0,1)+1$ & $t_4+1-0.25$ & $N(0,1)+2$ & $t_4+2-0.5$ \\
         $B3$ & $N(0,1)$ & $0.5N(-5,1)+0.5N(5,1)$ &  $N(0,1)+1$ & $0.5N(-5,1)+0.5N(5,1)+1-1$ & $N(0,1)+2$ & $0.5N(-5,1)+0.5N(5,1)+2-2$ \\
          $B4$ & $U(-2,2)$ & $t_4$ &  $U(-2,2)+1$ & $t_4+1-0.25$ & $U(-2,2)+2$ & $t_4+2-0.5$ \\
          $B5$ & $U(-2,2)$ & $0.5N(-5,1)+0.5N(5,1)$ &  $U(-2,2)+1$ & $0.5N(-5,1)+0.5N(5,1)+1-1$ & $U(-2,2)+2$ & $0.5N(-5,1)+0.5N(5,1)+2-2$ \\
          $B6$ & $t_4$ & $0.5N(-5,1)+0.5N(5,1)$ &  $t_4+1$ & $0.5N(-5,1)+0.5N(5,1)+1-1$ & $t_4+2$ & $0.5N(-5,1)+0.5N(5,1)+2-2$ \\
        \hline
         $C1$ & $N(0,1)$ & $N(0,1)-1$ &  $U(-2,2)$ & $U(-2,2)-1.25$ & $t_4$ & $t_4-1.5$ \\
          $C2$ & $N(0,1)$ & $N(0,1)-1$ &  $U(-2,2)$ & $U(-2,2)-1.5$ & $0.5N(-5,1)+0.5N(5,1)$ & $0.5N(-5,1)+0.5N(5,1)-2$ \\
          $C3$ & $N(0,1)$ & $N(0,1)-1$ &  $t_4$ & $t_4-1.5$ & $0.5N(-5,1)+0.5N(5,1)$ & $0.5N(-5,1)+0.5N(5,1)-2$ \\
          $C4$ & $U(-2,2)$ & $U(-2,2)-1$ &  $t_4$ & $t_4-1.5$ & $0.5N(-5,1)+0.5N(5,1)$ & $0.5N(-5,1)+0.5N(5,1)-2$ \\
        \hline
    \end{tabular}}
    \caption{Simulation scenarios}
\end{table}

\begin{table}[H]
    \centering
    \scalebox{0.75}{
    \begin{tabular}{|c|cc|cc|cc|cc|cc|cc|}
 \hline
$(n_1^\omega = n_1,n_2^\omega = n_2,n_2^\omega = n_3)$ & \multicolumn{2}{c|}{(10,10,10)} & \multicolumn{2}{c|}{(50,50,50)} & \multicolumn{2}{c|}{(100,100,100)} & \multicolumn{2}{c|}{(500,500,500)} &  
\multicolumn{2}{c|}{(50,100,150)} &
\multicolumn{2}{c|}{(150,100,50)}\\
 \hline
 Test & U test & LRT &  U test & LRT &  U test & LRT &  U test & LRT &  U test & LRT  &  U test & LRT \\
 \hline
 Scenario \\
         \hline
         $A1$ & 0.075 & 0.068 & 0.048 & 0.045 & 0.048 & 0.044 & 0.053 & 0.052 & 0.054 & 0.048 & 0.054 & 0.051\\
         $A2$ & 0.071 & 0.07 & 0.048 & 0.046 & 0.055 & 0.054 & 0.051 & 0.051 & 0.054 & 0.06 & 0.049 & 0.049 \\
         $A3$ & 0.078 & 0.065 & 0.056 & 0.059 & 0.056 & 0.051 & 0.051 & 0.056 & 0.064 & 0.055 & 0.052 & 0.046\\
         $A4$ &0.058 & 0.052 & 0.044 & 0.045 & 0.051 & 0.047 & 0.049 & 0.05 & 0.052 & 0.051 & 0.053 & 0.056\\
         $A5$ &0.067 & 0.064 & 0.06 & 0.056 & 0.056 & 0.056 & 0.052 & 0.053 & 0.052 & 0.051 & 0.054 & 0.054\\
         $A6$ & 0.07 & 0.064 & 0.045 & 0.044 & 0.057 & 0.054 & 0.046 & 0.044 & 0.052 & 0.054 & 0.048 & 0.046\\
         $A7$ & 0.074 & 0.073 & 0.049 & 0.052 & 0.047 & 0.051 & 0.06 & 0.056 & 0.055 & 0.052 & 0.048 & 0.048 \\
         \hline
          $B1$ & 0.082 & 0.076 & 0.05 & 0.052 & 0.051 & 0.049 & 0.052 & 0.055 & 0.057 & 0.053 & 0.061 & 0.062 \\
          $B2$ & 0.069 & 0.066 & 0.049 & 0.048 & 0.046 & 0.044 & 0.044 & 0.049 & 0.058 & 0.05 & 0.053 & 0.046  \\
          $B3$ & 0.07 & 0.089 & 0.042 & 0.052 & 0.044 & 0.044 & 0.04 & 0.046 & 0.057 & 0.06 & 0.053 & 0.048 \\
          $B4$ & 0.082 & 0.074 & 0.054 & 0.051 & 0.057 & 0.056 & 0.059 & 0.058 & 0.052 & 0.048 & 0.05 & 0.054\\
          $B5$ & 0.079 & 0.09 & 0.056 & 0.052 & 0.049 & 0.052 & 0.06 & 0.053 & 0.052 & 0.052 & 0.054 & 0.047 \\
          $B6$ & 0.066 & 0.079 & 0.056 & 0.056 & 0.044 & 0.048 & 0.043 & 0.048 & 0.052 & 0.054 & 0.053 & 0.059 \\
\hline
 $C1$ & 0.076 & 0.07 & 0.06 & 0.06 & 0.046 & 0.044 & 0.052 & 0.055 & 0.052 & 0.046 & 0.054 & 0.052\\
          $C2$ &0.072 & 0.073 & 0.062 & 0.062 & 0.051 & 0.05 & 0.054 & 0.054 & 0.05 & 0.048 & 0.052 & 0.05\\
          $C3$ & 0.066 & 0.073 & 0.066 & 0.067 & 0.05 & 0.053 & 0.051 & 0.045 & 0.052 & 0.049 & 0.072 & 0.066\\
          $C4$ & 0.064 & 0.06 & 0.044 & 0.05 & 0.052 & 0.052 & 0.054 & 0.054 & 0.059 & 0.059 & 0.052 & 0.055\\
         \hline
    \end{tabular}}
    \caption{Rejection rates of null cases under various settings}
\end{table}

\begin{figure}[H]
  \centering
  \begin{tabular}[b]{c}
    \includegraphics[width=7in]{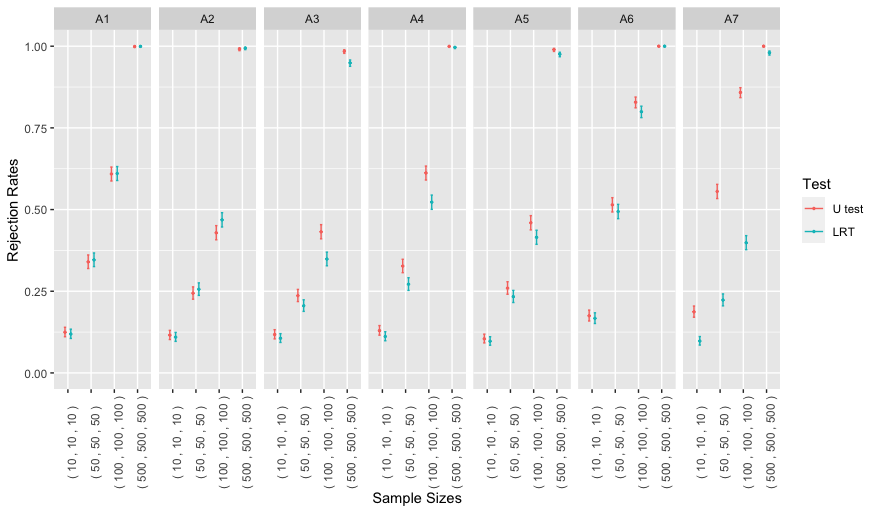} \\
    \small (a) Cases where sample sizes are consistent across strata
  \end{tabular} \qquad
  \vspace{0.1in}
  \begin{tabular}[b]{c}
    \includegraphics[width=7in]{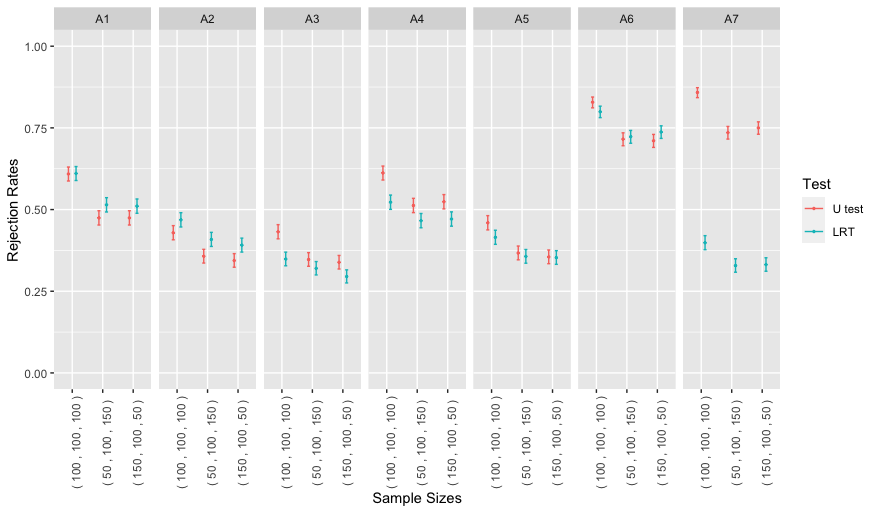} \\
    \small (b) Cases where sample sizes can change across strata
  \end{tabular}
  \caption{Rejection rates and their 95\% confidence intervals of alternative cases in \textit{Case A}}
\end{figure}

\begin{figure}[H]
  \centering
  \begin{tabular}[b]{c}
    \includegraphics[width=7in]{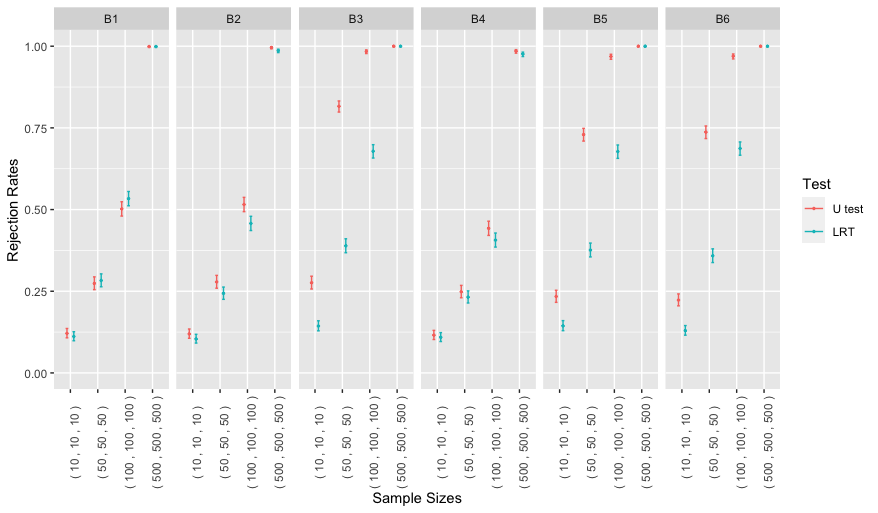} \\
    \small (a) Cases where sample sizes are consistent across strata
  \end{tabular} \qquad
  \vspace{0.1in}
  \begin{tabular}[b]{c}
    \includegraphics[width=7in]{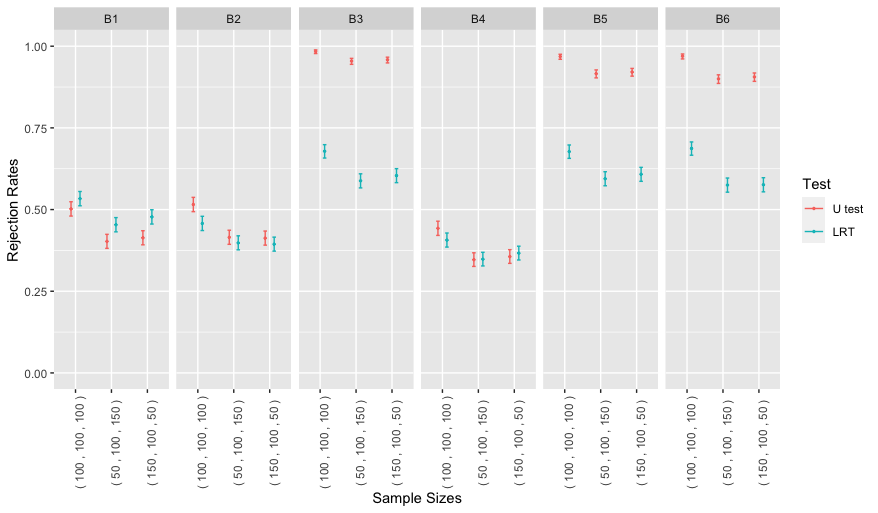} \\
    \small (b) Cases where sample sizes can change across strata
  \end{tabular}
  \caption{Rejection rates and their 95\% confidence intervals of alternative cases in \textit{Case B}}
\end{figure}

\begin{figure}[H]
  \centering
  \begin{tabular}[b]{c}
    \includegraphics[width=7in]{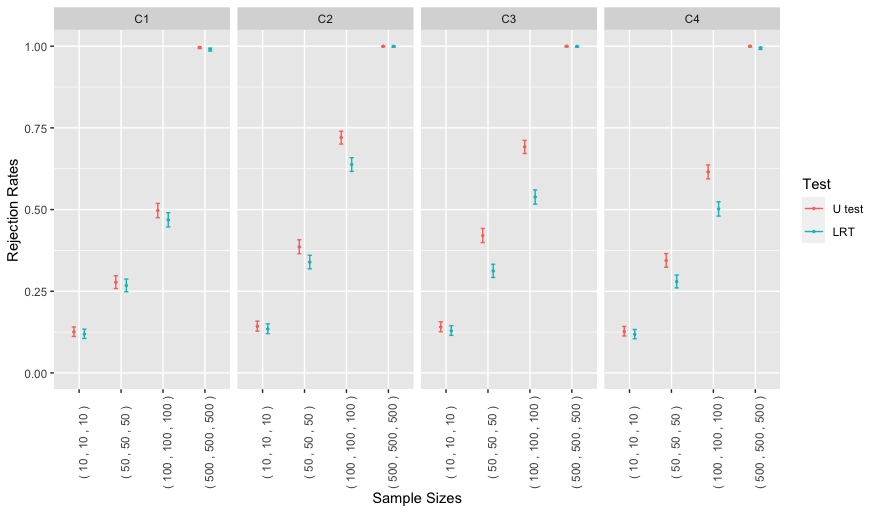} \\
    \small (a) Cases where sample sizes are consistent across strata
  \end{tabular} \qquad
  \vspace{0.1in}
  \begin{tabular}[b]{c}
    \includegraphics[width=7in]{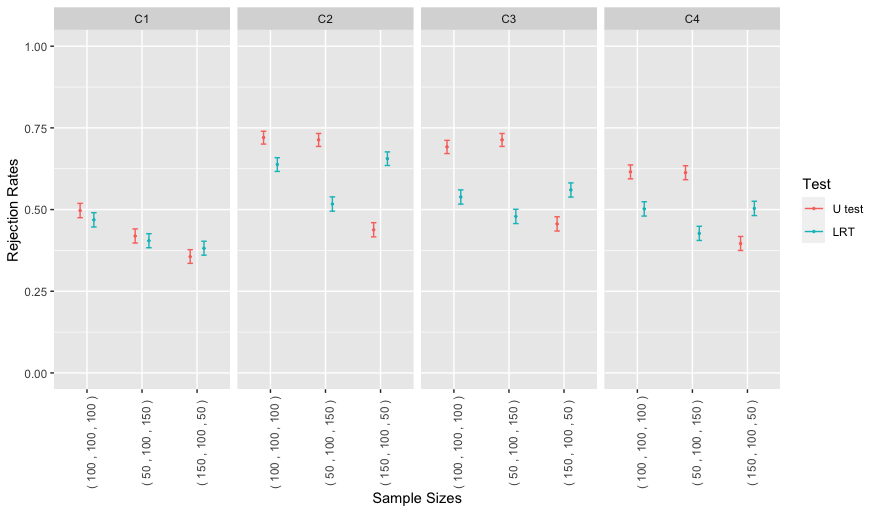} \\
    \small (b) Cases where sample sizes can change across strata
  \end{tabular}
  \caption{Rejection rates and their 95\% confidence intervals of alternative cases in \textit{Case C}}
\end{figure}

\subsection{Investigating the Power of the Tests}
The results in Section 4.4 focus on only a single non-null example for each scenario. this section investigate the power as a function of sample size for different treatment effects. We use the scenarios $A3$ and $A4$, and carried out simulations as described in the previous section.

For scenario $A3$, we generated three strata with $ n_s^\omega = n ~( s\in \{1,2,3\}, \omega \in \{t,c\})$, and generated random samples from $t_4$ distribution with location shifts comprised of strata effects $\Delta_1=0, \Delta_2=1, \Delta_3=2$ and additive treatment effects $\tau_1 = 1$, $\tau_2=1+\Gamma$ and $\tau_3 = 1+2\Gamma$.
Here the sequence of the treatment effects $\{\tau_1, \tau_2, \tau_3\}$ is arithmetic and we treat $\Gamma$ as the effect size. 
For each fixed effect size, we explore the relationship between the sample size $n$ and the rejection rates for both tests, and the results are shown in Figure 4 with $\Gamma$ ranging from 0 to 0.5, and $n$ ranging from 10 to 1000. In alternative cases when $\Gamma>0$, with each fixed $\Gamma$, as $n$ increases, the rejection rates of both tests increase, and the power of the U test is always higher than the LRT for each $n$.

\begin{figure}[H]
    \centering
    \includegraphics[width = 3.5in]{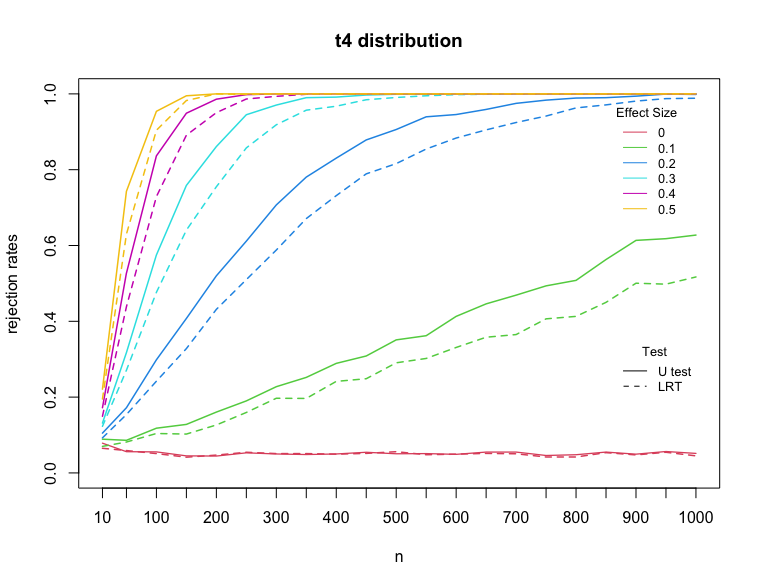}
    \caption{Relationship between rejection rates and sample sizes for $t_4$ distribution}
\end{figure}

For scenario $A4$, again we generated three strata with $ n_s^\omega = n ~( s\in \{1,2,3\}, \omega \in \{t,c\})$. With 
$\frac{Y_s^\omega}{c_s^\omega} \sim \chi_1^2$, we took the logarithm of $Y_s^\omega$ as the outcome. So the treatment effect is defined as $\tau_s = log(c_s^t) - log(c_s^c)$. We focus on the case where the sequence of the treatment effects $\{\tau_1,\tau_2,\tau_3\}$ are arithmetic with $\tau_2 = \tau_1+\Gamma$ and $\tau_3 = \tau_1+2\Gamma$, and $\Gamma$ is the effect size. By fixing $ log(c_1^c) = 0,log(c_2^c) = 1, log(c_3^c) = 2$ and $log(c_1^t) = 1$, we have $\tau_1 = 1$. Then we can get different values of $\Gamma$ by changing the values of $c_2^t$ and $c_3^t$. Then for each $\Gamma$, we can explore the relationship between the rejection rates of our U test and the LRT as the sample size $n$ varies. Figure 5 shows the rejection rates with $n$ ranging from 10 to 1000 when $\Gamma$ ranging from 0 to 1. As we would expect, with fixed $n$, larger effect size leads to larger rejection rates. The rejection rates of the U test is always larger than the LRT for each fixed $n$ and $\Gamma$.
\begin{figure}[H]
    \centering
    \includegraphics[width = 3.5in]{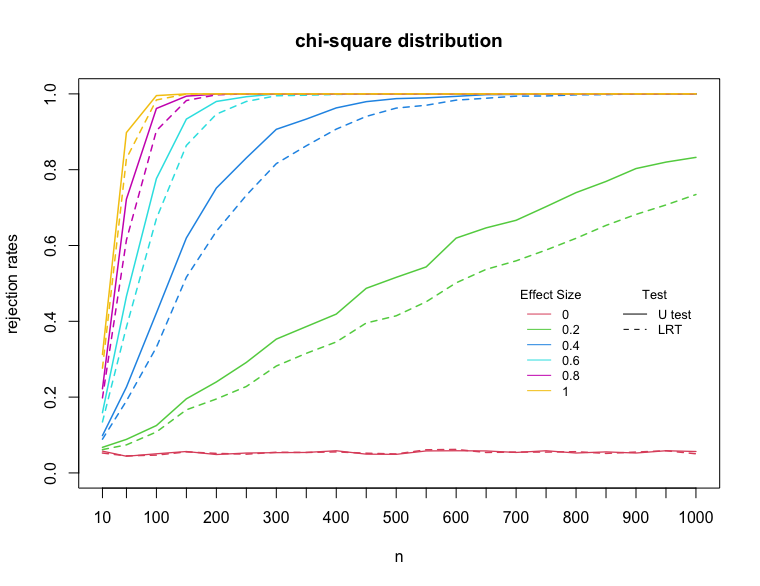}
    \caption{Relationship between rejection rates and sample sizes for $\chi_1^2$ distribution}
\end{figure}

\section{Case Study}

In this section, we apply our proposed U test to a randomized data set from a program evaluation study in labor economics, an evaluation of the National Supported Work (NSW) Demonstration. The NSW is a labor training program conducted in the mid-1970s aiming at providing work experience to people with economic difficulties. Please refer to \cite{lalonde1986evaluating} and \cite{dehejia1999causal} for details about the program. We use a subset of the \cite{lalonde1986evaluating} data that was created and used by \cite{dehejia1999causal}. These data described results from male participants with earnings information available for 1974. Earnings in 1978 were treated as outcome, and several pretreatment variables were recorded. There are 185 subjects in treatment group and 260 subjects in control group.

In this randomized study, pretreatment variables should have the same distribution between the treatment and control groups. So we can directly compare the distributions of the outcome, 1978 earnings, for the treatment and control groups to get the treatment effect. The outcome distributions of the treatment and control groups are shown in Figure 6. Both of them are heavily right-skewed and have an excess of 0 values. Because the distributions are far from a normal distribution, a non-parametric test is more appropriate than parametric test assuming normality. The p-value of Mann-Whitney test is 0.01,
and the test statistic is 0.43. Here the expectation of the test statistic is the probability that a random outcome in the treatment group is smaller than a random outcome in the control group. The result indicates that there is a positive treatment effect.

Next we construct strata based on two important pretreatment variables, age and 1974 earnings, separately, and then apply our proposed U test to identify whether there is treatment effect heterogeneity across the strata. We first split all subjects by quartiles of age. The subgroup sample sizes are in Table 3, and the pairwise U-statistics are in Table 4. Here the expectation of $U^{(p,q)}$ is the probability that the difference between treatment and control outcomes in stratum $p$ are smaller than the difference in stratum $q$. As the $U$ values are greater then 0.5, the treatment effects in younger strata are generally smaller than those in older strata. However, the p-value of our proposed heterogeneity test is 0.58, so the observed heterogeneity is not statistically significant.

Then we explore whether the treatment effect differs between participants with and without positive incomes in 1974. The first stratum is for participants without income in 1974 and the second stratum is for those with positive income. The subgroup sample sizes are shown in Table 5. The U-statistic comparing their treatment effects $U^{(1,2)}$ is 0.409, and the p-value of our heterogeneity test is 0.032, which indicates this program has greater impact for participants who did not have any income in 1974 than those who had some income. 

\begin{figure}[H]
    \centering
    \includegraphics[width = 3.5in]{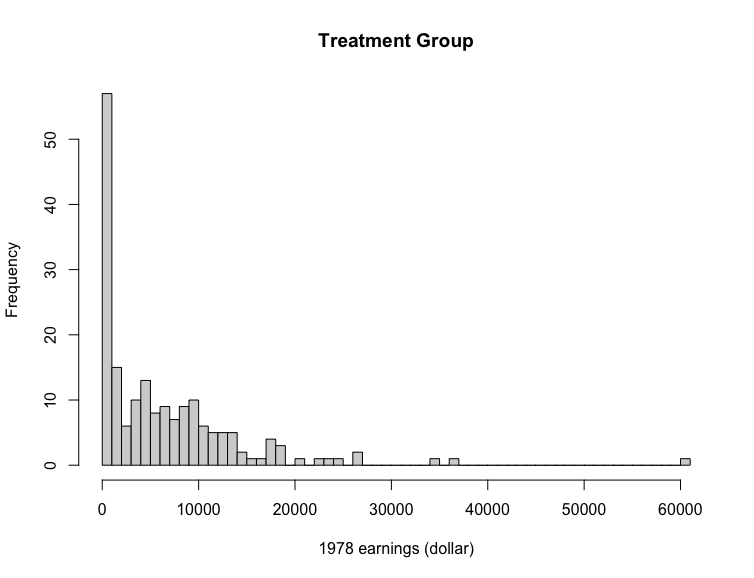}
    \includegraphics[width = 3.5in]{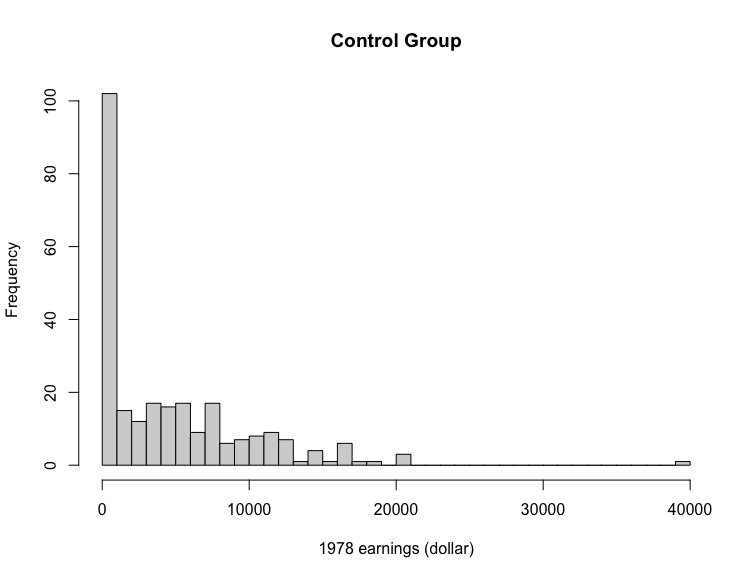}
    \caption{Distribution of 1978 earnings in the treatment and control groups}
\end{figure}

\begin{table}[H]
\centering
\begin{tabular}{rrrrr}
  \hline
  Stratum & 1 & 2 & 3 & 4\\
 Age & [17, 20] & (20,24] & (24,28] & (28, 55] \\ 
  \hline
Treatment &  47 &  41 &  49 &  48 \\ 
  Control &  83 &  56 &  60 &  61 \\ 
   \hline
\end{tabular}
\caption{Sample sizes in different treatment and age groups}
\end{table}

\begin{table}[H]
\centering
\begin{tabular}{rrrrrrrr}
  \hline
 & $U^{(1,2)}$ & $U^{(1,3)}$ & $U^{(1,4)}$ & $U^{(2,3)}$ & $U^{(2,4)}$ & $U^{(3,4)}$\\ 
  \hline
 & 0.52 & 0.55 & 0.57 & 0.53 & 0.55 & 0.51\\ 
   \hline
\end{tabular}
\caption{Pairwise U-statistics}
\end{table}

\begin{table}[H]
\centering
\begin{tabular}{rrr}
  \hline
  Stratum & 1 & 2\\
1974 Income & 1974 Income = 0 & 1974 Income $>$ 0 \\ 
  \hline
Treatment & 131 &  54 \\ 
  Control & 195 &  65 \\ 
   \hline
\end{tabular}
\caption{Sample sizes of different treatment and income groups}
\end{table}


\section{Discussion}
Identifying the existence of treatment effect heterogeneity is a key element of attempts to provide more precise treatment recommendations for individuals. We have described a U-statistic-based approach to formally test the hypothesis of homogeneous treatment effects without assuming a particular parametric form of the outcome distributions, and compared its performance with the LRT. The LRT requires the distribution of treatment effect estimates to be normal, which can be satisfied if the outcomes are normal or the sample sizes are large (by the Central Limit Theorem). Our results show that, as expected, when the outcome distributions are close to normal, the power of the LRT is a little better than the U test. However when at least one of the outcome distributions departs substantially from the normal distribution, the power of our non-parametric test can be significantly larger than the LRT. As the departure increases, the advantage of the U test increases. This observation is similar to the comparison between Mann-Whitney test and t-test \citep{lehmann2004elements}.

Though our approach can be applied to any situations as long as the two assumptions addressed in the second paragraph of Section 3 are satisfied, we do need to be cautious about understanding what aspect of the distribution is being compared by the U test. The U test is most easily interpreted when we believe that one of the scenarios described in Section 3.3 applies. For those scenarios (\textit{Case A}, \textit{B} and \textit{C}), the test is comparing locations of distributions assumed to come from common families.

One limitation of this method is that it requires the distributions of all confounding variables are the same between the treatment group and the control group within stratum, which is true in randomized experiments. However, in observational studies, we will need to adjust for confounding variables. Even if the strata are created based on estimated propensity scores in an effort to balance the baseline covariates \citep{xie2012estimating}, some further adjustments for remained imbalance may be needed.

\section*{}
\bibliography{bibliography3}

\end{document}